\documentclass[fleqn,usenatbib]{mnras}

\usepackage[T1]{fontenc}

\DeclareRobustCommand{\VAN}[3]{#2}
\let\VANthebibliography\thebibliography
\def\thebibliography{\DeclareRobustCommand{\VAN}[3]{##3}\VANthebibliography}

\usepackage{graphicx}	
\usepackage{amsmath}	
\usepackage{amssymb}	

\usepackage{graphicx}
\newcommand\U{H}
\newcommand\F{F}
\newcommand\VR{VR}
\newcommand\EQ{EQ}
\newcommand\tECC{}
\newcommand\uECC{uE}
\newcommand\Coll{}
\newcommand\nColl{nC}

\newcommand\UVR{\U\VR\Coll\tECC}
\newcommand\DVR{\F\VR\Coll\tECC}
\newcommand\UEQ{\U\EQ\Coll\tECC}
\newcommand\DEQ{\F\EQ\Coll\tECC}
\newcommand\UuECCVR{\U\VR\Coll\uECC}
\newcommand\DuECCVR{\F\VR\Coll\uECC}
\newcommand\UuECCEQ{\U\EQ\Coll\uECC}
\newcommand\DuECCEQ{\F\EQ\Coll\uECC}
\newcommand\UnCVR{\U\VR\nColl\tECC}

\newcommand\UnCREQ{\U\EQ\nColl\tECC}

\usepackage{newtxtext,newtxmath}
\title[Binary survival in UFDs]{Evolution of binary stars in the early evolutionary phases of ultra-faint dwarf galaxies}
\author[A. R. Livernois et al.]{Alexander R. Livernois,$^{1}$\thanks{E-mail: allivern@iu.edu}
Enrico Vesperini,$^{1}$
V\'aclav Pavl\'ik$^{1,2}$
\\
$^{1}$Department of Astronomy, Indiana University, Bloomington, IN, 47405, USA\\
$^{2}$Department of Physics, Indiana University, Bloomington, IN, 47405, USA\\
}

\begin{document}

\label{firstpage}
\pagerange{\pageref{firstpage}--\pageref{lastpage}}
\maketitle

\begin{abstract}
The dynamics of binary stars provides a unique avenue to gather insight into the study of the structure and dynamics of star clusters and galaxies. 
In this paper, we present the results of a set of $N$-body simulations aimed at exploring the evolution of binary stars during the early evolutionary phases of ultra-faint dwarf galaxies (UFD). 
In our simulations, we assume that the stellar component of the UFD is initially dynamically cold and evolves towards its final equilibrium after undergoing the violent relaxation phase. 
We show that the early evolutionary phases of the UFD significantly enhance the disruption of wide binaries and leave their dynamical fingerprints on the semi-major axis distribution of the surviving binaries as compared to models initially in equilibrium. 
An initially thermal eccentricity distribution is preserved except for the widest binaries for which it evolves towards a superthermal distribution; for a binary population with an initially uniform eccentricity distribution, memory of this initial distribution is rapidly lost for most binaries as wider binaries evolve to approach a thermal/superthermal distribution. 
The evolution of binaries is driven both by tidal effects due to the potential of the UFD dark matter halo and collisional effects associated to binary-binary/single star encounters. 
Collisional effects are particularly important within the clumpy substructure characterizing the system during its early evolution; in addition to enhancing binary ionization and evolution of the binary orbital parameters, encounters may lead to exchanges of either of the primordial binary components with one of the interacting stars.

\end{abstract}


\section{Introduction}
Wide binary stars provide a unique and powerful tool  for the investigation of the dynamical effects associated with the gravitational potential of their host stellar systems and their interactions with other stars. Due to their low binding energies and high cross-sections, they are sensitive to destruction and mutation from gravitational tidal fields from galactic to molecular cloud scales and interactions with individual passing stars (e.g. \citealt{1944Ch}, \citealt{1975He}, \citealt{1987WeSh}, \citealt{2010JiTr}, \citealt{2016PeLu}). Thus, the distribution functions of binary internal parameters, such as semi-major axis and eccentricity, contain a wealth of fundamental information concerning not only their formation history, but also the dynamical environment they live in (e.g. \citealt{2019HaRa}, \citealt{2019HaRa2}, \citealt{2019GeLe}, \citealt{2022HaCh}, \citealt{2021PeJo}, \citealt{2023BeEv}). 

Data from the Gaia mission have enabled the study of these systems on a large scale (see e.g. \citealt{2016Gaia}, \citealt{2017OhPr}, \citealt{2018ElRi}, \citealt{2020HaLe}, \citealt{2020TiEl}, \citealt{2021ElRi}) and provided new opportunities for extensive studies of the distribution of their internal properties in the solar neighborhood (see e.g. \citealt {2016ToKi}, \citealt{2020Toko}, \citealt{2020TiEl}, \citealt{2022HwTi}, \citealt{2022HwEl}, \citealt{2022RaBu}). 

Binaries have also received attention when studying ultra-faint dwarf galaxies; in particular, a number of investigations have explored how they may bias the observational estimates of the velocity dispersion of these systems (see e.g. \citealt{2010McCo}, \citealt{2017SpMa}, \citealt{2019MiPa}; see also  \citealt{2019Si} and references therein) and how their evolution and survival may add further insight into the mass distribution of their host dark matter halos and the core vs cusp problem (\citealt{2016PeLu}, \citealt{2022RaBu}). While the direct observation of wide binary pairs in these systems is difficult, some studies have utilized the two-point correlation function to ascertain a distribution function of their semi-major axis (\citealt{2022KeWa}, \citealt{2022SaSi}).

In this paper, we study the disruption and evolution of wide binary stars during the early evolutionary phases of ultra-faint dwarf galaxies. We focus our attention on the dynamically cold collapse and violent relaxation characterizing the early formation  and evolutionary phases of dwarf galaxies (e.g. \citealt{2016RiPa}, \citealt{2020LaNa}); in particular, we investigate how the early collapse and evolution towards an equilibrium configuration affect the strength of the tidal effects on primordial binaries and how the structural evolution during these phases implies that the collisional effects associated with stellar encounters (which are usually not important and therefore neglected in the study of the dynamics of ultra-faint dwarfs) can play a significant role in the evolution of binary stars.   

This paper is structured as follows: in Section \ref{sec:Methods}, we introduce the models and numerical simulations we use to study ultra-faint dwarf systems; in Section \ref{sec:Results}, we investigate the survival and evolution of internal properties of our binary systems; and finally in Section \ref{sec:Concl}, we summarize our findings.

\begin{table}

\centering
\begin{tabular}{|l|l|l|l|}
\hline
ID.  & Initial Struct.& Integration & Ecc. Dist.  \\ \hline
\UVR\ & Homogeneous & \textsc{NBODY6++GPU} & Thermal        \\
\UuECCVR\ & Homogeneous & \textsc{NBODY6++GPU} & Uniform \\ 
\UnCVR& Homogeneous  & Collisionless & Thermal  \\
\UEQ\   & Equilibrium (\UVR) & \textsc{NBODY6++GPU}& Thermal    \\ 
\UuECCEQ\ & Equilibrium (\UVR)& \textsc{NBODY6++GPU} & Uniform   \\
\UnCREQ\ & Equilibrium (\UVR) & Collisionless  & Thermal  \\
\DVR  & Fractal (D=2)    & \textsc{NBODY6++GPU} & Thermal    \\
\DuECCVR\ & Fractal (D=2)    & \textsc{NBODY6++GPU}& Uniform\\
\DEQ\   & Equilibrium (\DVR) & \textsc{NBODY6++GPU} & Thermal\\
\DuECCEQ\ & Equilibrium (\DVR)   & \textsc{NBODY6++GPU} & Uniform  \\
\hline
\end{tabular}

\caption{Summary of initial conditions for the simulations presented in this paper. All models have 12,500 binary pairs generated following the eccentricity distribution in the table. The violent relaxation models have extra binaries at time 0 due to coincidental random spacings and motions. The equilibrium models are generated from the respective violent relaxation models at  $\approx 30t_{\rm ff}$.}
\label{SimTable}
\end{table}

\section{Initial Conditions and Integration}
\label{sec:Methods}
In this study, we integrate 8 models using the GPU-accelerated \textsc{NBODY6++GPU} code \citep{2015WaRa}.
All of our initial conditions have $N=5\times10^4$ equal-mass stars with mass $m_*$. Half of the stars are generated in binary pairs with a log-uniform semi-major axis distribution and an eccentricity distribution following a power-law distribution:
\begin{equation}
    P(e)\propto e^{\alpha}
    \label{eq:ecc}
\end{equation}
\noindent with either $\alpha=0$ for our uniform eccentricity models (denoted \uECC) or $\alpha=1$ for our thermal eccentricity models (no denotation).

The stellar systems we study are initially placed at the center of  a static dark matter potential following the Navarro-Frenk-White profile (NFW; \citealt{1997NFW}) defined here as:
\begin{equation}
 \phi = \frac{GM_{\rm s}}{r} {\rm ln} (1+r/R_{\rm s})    
\label{eq:NFWPot}
\end{equation}

\noindent where $M_{\rm s}$ and $R_{\rm s}$ are the scale mass and scale radius, respectively.

We have explored the following initial conditions.
A subset of models start with  dynamically cold systems (denoted \VR) with a virial ratio of $Q_*=10^{-3}$, where $Q_*$ is defined here as the ratio of the total kinetic energy of the stellar system to its potential energy (calculated including only the stellar component). 
For these simulations, we have considered systems with either a homogeneous spatial distribution (denoted \U) or a fractal spatial distribution with fractal dimension $D=2$ (denoted \F) with half mass radii $\approx$ 0.18 $R_{\rm s}$. The ratio of the median local density of the fractal model (calculated using the sixth nearest neighbour; see \citealt{CasHut}) to that of the homogeneous model is equal to about 40. These initial conditions were generated using the \textsc{McLuster} code \citep{2011KuMa}.
The main goal of this paper is to explore the evolution of the binary star population during the early evolution and violent relaxation phase of these dynamically cold systems.

In order to clearly illustrate the role of these early dynamical phases, we have compared the results of these simulations with those of another set of models starting in equilibrium with positions and velocities 
equal to those reached by the corresponding \VR\ models after 30 $t_{\rm ff}$ (denoted \EQ), where $t_{\rm ff}$ is the free-fall timescale and is defined below.
In these equilibrium initial conditions, the population of primordial binaries has been set with the same initial properties (number, eccentricity and semi-major axis distribution) as those adopted in the initial conditions of the corresponding \VR\ models.

Two of our models (\UVR, \UEQ) have also been run using a code that integrates the evolution of each binary star independently in the same dark matter halo potential and starting from the same initial conditions but without considering the collisional effects of encounters with other single stars, binary stars, or the effect of the stellar component of the system's potential (denoted \nColl).  
These additional integrations allow us to disentangle the effects associated with the tidal field of the dark matter halo (the only ones affecting binaries in these integrations) from those due to stellar interactions which affect binaries in the full, self-consistent $N$-body simulations run with \textsc{NBODY6++GPU}.

The IDs used to identify the various models along with their initial properties are listed in Table \ref{SimTable}.

We point out that the population of binaries in our initial conditions may be altered by two effects: some binary pairs do not survive once all stars are added in the $N$-body system since a single star, either component of another binary pair, or another whole binary pair may supplant one of the original members in the binary system, and some pairs of single stars may be bound in binary pairs. These effects happen most frequently in our dynamically cold initial conditions and in the wide binary pair population. 

In a few cases, in order to directly compare the impact of the evolution of stellar systems with different initial structural and kinematic properties on binary stars, we use (for each of the eccentricity distribution considered) a set $\mathcal{I}$ of binaries composed of binaries that we generate with the preset internal properties described above and survive the introduction of single stars and other binaries into the system in all models.

In our dynamically cold models, our systems collapse in a range of timescales shorter than the free-fall time associated with the stellar system alone due to the potential of the dark matter halo in which these systems are embedded. As a reference timescale for our analysis, we adopt the time needed by the stellar system to reach the smallest half-mass radius during its initial collapse; hereafter we refer to this time as $t_{\rm ff}$. We point out that  throughout the paper, mentions of the semi-major axis and eccentricity of orbits refer to the properties of the inner orbits of binary systems;  all other terms (such a energy, angular momentum, radius, etc.) refer to the outer orbit of the single stars, or center of mass for binary pairs, in the potential of the total stellar and DM system.

Finally, as a reference physical scaling for our models we adopt the following values:  $m_*=0.5\ M_{\sun}$, $M_{\rm s}=10^8\ M_{\sun}$, $R_{\rm s}=500$ pc, $t_{\rm ff}=16.1$ Myr; the log-uniform distribution of semi-major axes of our generated binaries is between $a_{\rm min} = 0.0002$ pc and $a_{\rm max}=2$ pc, and the simulations are ran out to 1 Gyr.

\section{Results}
\label{sec:Results}
\subsection{A preliminary estimate of binary star disruption}
A preliminary estimate of the survival of binary stars in an external potential can be calculated following the framework outlined, for example, in \cite{2016PeLu}. This calculation is based on the estimate of the tidal radius of binaries along their orbits in the dark matter potential (for this calculation only this potential is considered while the potential of the stellar component is neglected). Specifically, the tidal radius, $r_{\rm t}$, is calculated as follows:

\begin{equation}
    r_{\rm t}= \left[\frac{Gm_{\rm b}}{\lambda_1}\right]^{1/3}
\label{eq:r1}
\end{equation}

\noindent where $G$ is the gravitational constant, $m_{\rm b}$ is the total mass of the binary pair, and:
\begin{equation}
    \lambda_1 \approx -\frac{d\rm{log}\rho}{d\rm{log}r} \Omega^2,
\label{eq:l1}
\end{equation}
where $r$ is the separation between the center of mass of the binary pair and the center of the NFW potential, $\Omega$ is the circular angular velocity of the potential, and $\rho$ is the mass density of the potential. Evaluating $\lambda_1$ for the NFW halo, we retrieve:
\begin{equation}
    \lambda_1 \approx \frac{GM_{\rm s}}{r^2}\left(\frac{2r/R_{\rm s}}{1+r/R_{\rm s}}+1\right)
    \left(\frac{{\rm ln}(1+r/R_{\rm s})}{r}-\frac{1}{r+R_{\rm s}}\right).
\label{eq:ltot}
\end{equation}

Utilizing equations \ref{eq:r1}-\ref{eq:ltot}, we derive the following expression for $r_{\rm t}$ as a function of $r$ for the binary pairs in our models:
\begin{equation}
\resizebox{0.45\textwidth}{!}{$
    r_{\rm t}(r) = \left[\frac{M_{\rm s}}{m_{\rm b}r^2}\left(\frac{2r/R_{\rm s}}{1+r/R_{\rm s}}+1\right)
    \left(\frac{{\rm ln}(1+r/R_{\rm s})}{r}-\frac{1}{r+R_{\rm s}}\right)\right]^{-1/3}.
    $}
\label{eq:rtfin}
\end{equation}

For each binary, we estimate the tidal radius at the pericenter of the binary's outer orbit in the dark matter potential, $r_{\rm t}(r_{\rm p})$, and if this radius is smaller than the binary's semi-major axis, $a$, the binary is considered disrupted.

\begin{figure}
    \centering
    \includegraphics[width=0.48\textwidth]{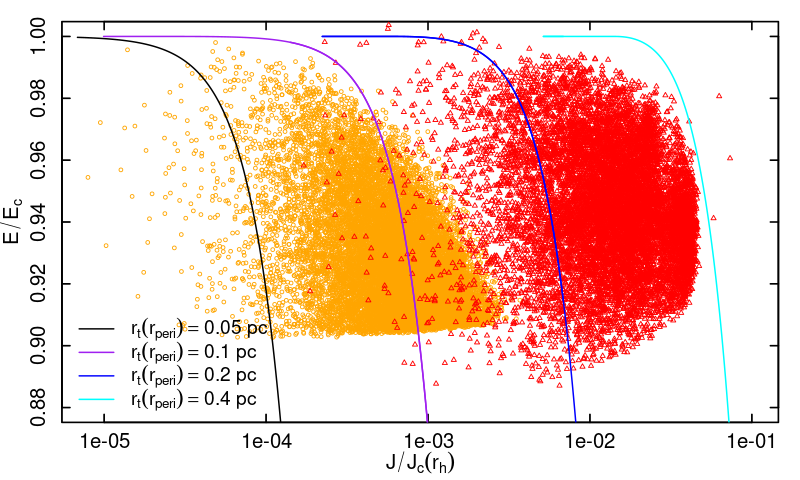}
    \caption{Initial outer $E$-$J$ distribution of all binary pairs in the \UVR\ (orange circles) and \UEQ\ (red triangles) models, normalized by the circular energy of the outer orbit, $E_{\rm c}$, and the circular specific angular momentum at the half-mass radius of the equilibrium system, $J_{\rm c}(r_{\rm h})$. Lines of constant minimum tidal radius (see equation \ref{eq:rtfin}) are plotted for reference of expected binary survival. Note the significant differences in the orbital initial conditions between the two models and the subsequent expected destruction of binary pairs.}
    \label{fig:EJstability}
\end{figure}

In Figure \ref{fig:EJstability}, we visualize this estimation of the tidal radius at the pericenter by plotting contours of constant $r_{\rm t}(r_{\rm p})$ on the $E$-$J$ plane and show the positions of all the binaries initially in  our \UVR\ and \UEQ\ models on this plane.

This figure provides a preliminary illustration of how the orbital properties of stars in a sub-virial cold stellar system lead to a more efficient disruption of binary stars than a system initially in equilibrium. We emphasize that in this very simple estimate, various additional dynamical processes affecting the evolution and disruption of binary stars are neglected. Specifically, the calculation presented above does not include the effects of the time variation of the tidal field of the external dark matter halo along the binary's orbit and the collisional effects associated with single-binary and binary-binary encounters. These encounters are more frequent within the clumpy substructure characterizing the system during its early evolution and when a system undergoing a violent relaxation collapse reaches its most compact configuration at the time of maximum contraction.

The $N$-body simulations run for this paper include all of these processes, and we explore their effects on binary evolution in the following subsections.

\begin{figure}
    \centering
    \includegraphics[width=0.45\textwidth]{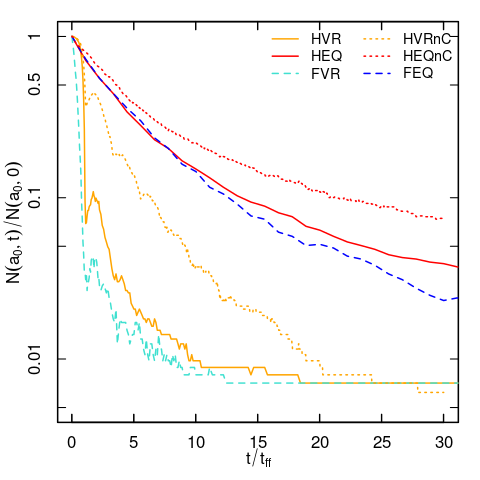}
    \caption{Time evolution of the fraction of the binaries in set $\mathcal{I}$ with $0.2<a_0<2$ pc, for a selection of models. Note the clear difference in destruction between the violent relaxation and equilibrium models, and the additional difference between the collisional and collisionless models.}
    \label{fig:Bincount}
\end{figure}

\begin{figure}
    \centering
    \includegraphics[width=0.4\textwidth]{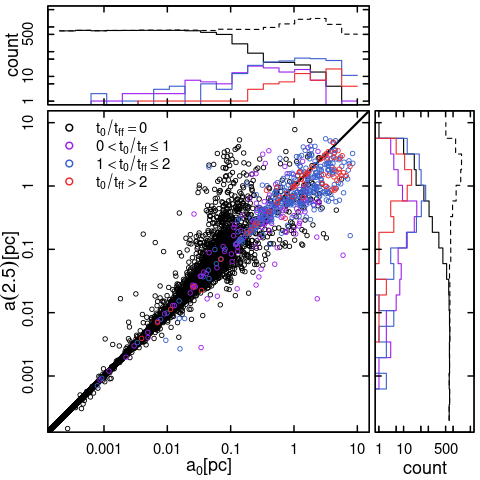}
    \includegraphics[width=0.4\textwidth]{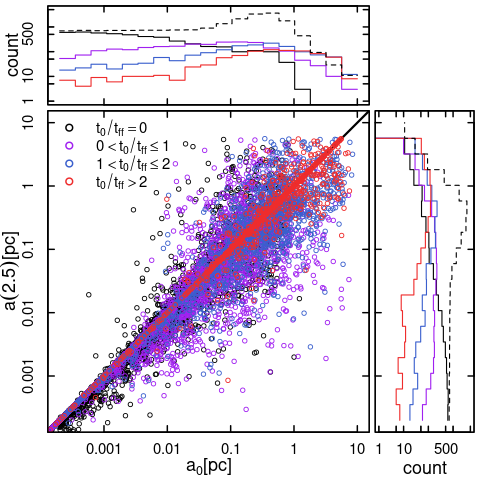}
    \includegraphics[width=0.4\textwidth]{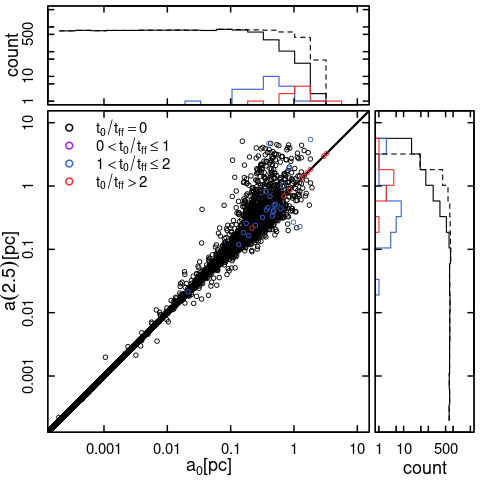}
    \caption{Distribution of binary pairs in $a(2.5)$ (semi-major axis at time t=2.5$t_{\rm ff}$), $a_0$ (semi-major axis when formed), and on the $a(2.5)$-$a_0$ plane for the \UVR, \DVR, and \UEQ\ models (top to bottom). Colors are determined by formation time for each binary ($t_0$). For each histogram, the initial distribution of binary semi-major axes at $t=0$ is plotted for reference as a dashed black line.}
    \label{fig:aarep}
\end{figure}
\begin{figure*}
    \centering
    \includegraphics[width=\textwidth]{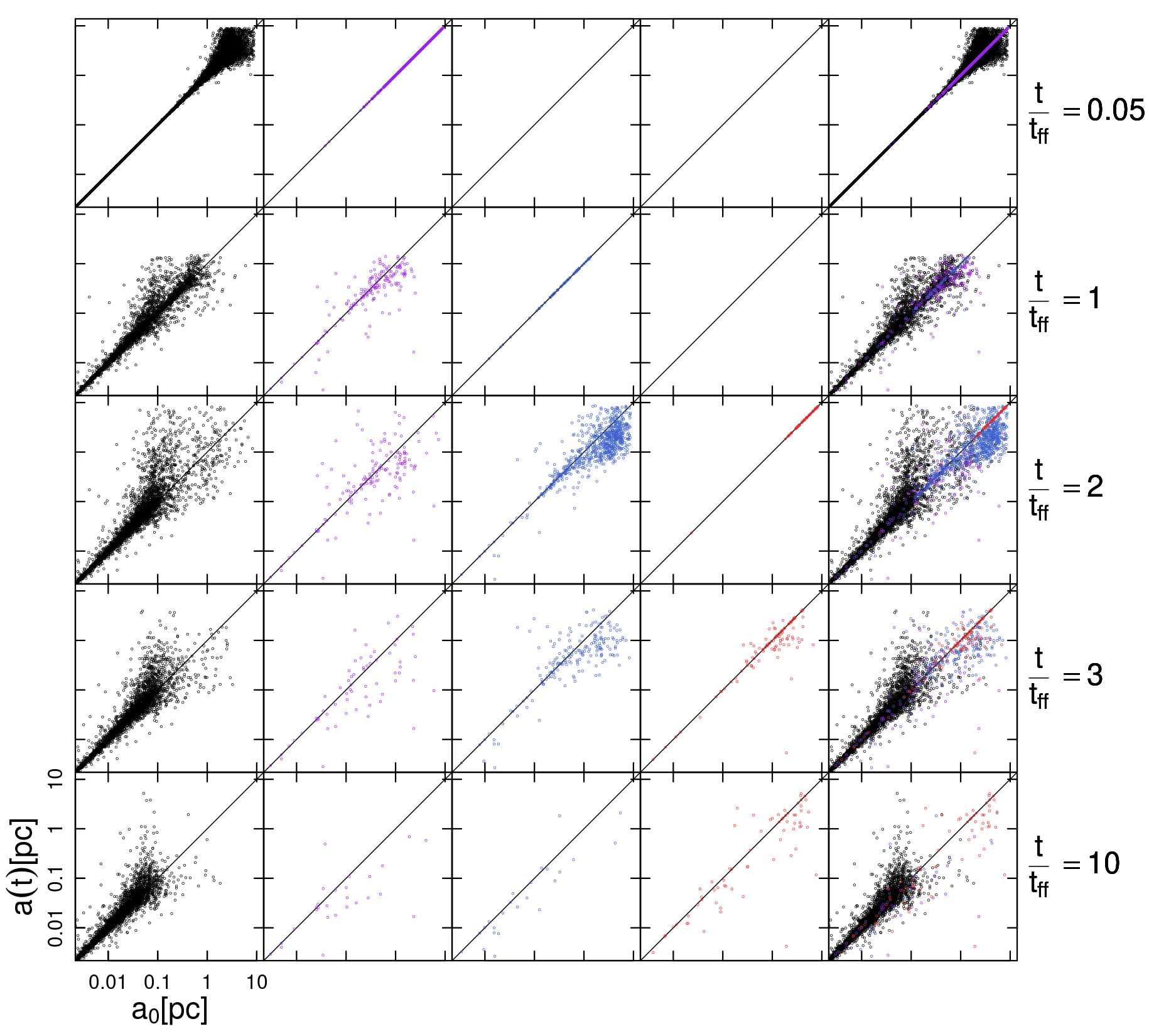}
    \caption{Distribution of binaries on the $a(t)$-$a_0$ plane for the \UVR\ model. Colors are determined by formation time for each binary ($t_0$): primordial (black), $0< t_0/t_{\rm ff}<1$ (purple), $1\leq t_0/t_{\rm ff}<2$ (blue), $t_0/t_{\rm ff}\geq2$ (red).}
    \label{fig:uvraa}
\end{figure*}

\begin{figure*}
    \centering
    \includegraphics[width=\textwidth]{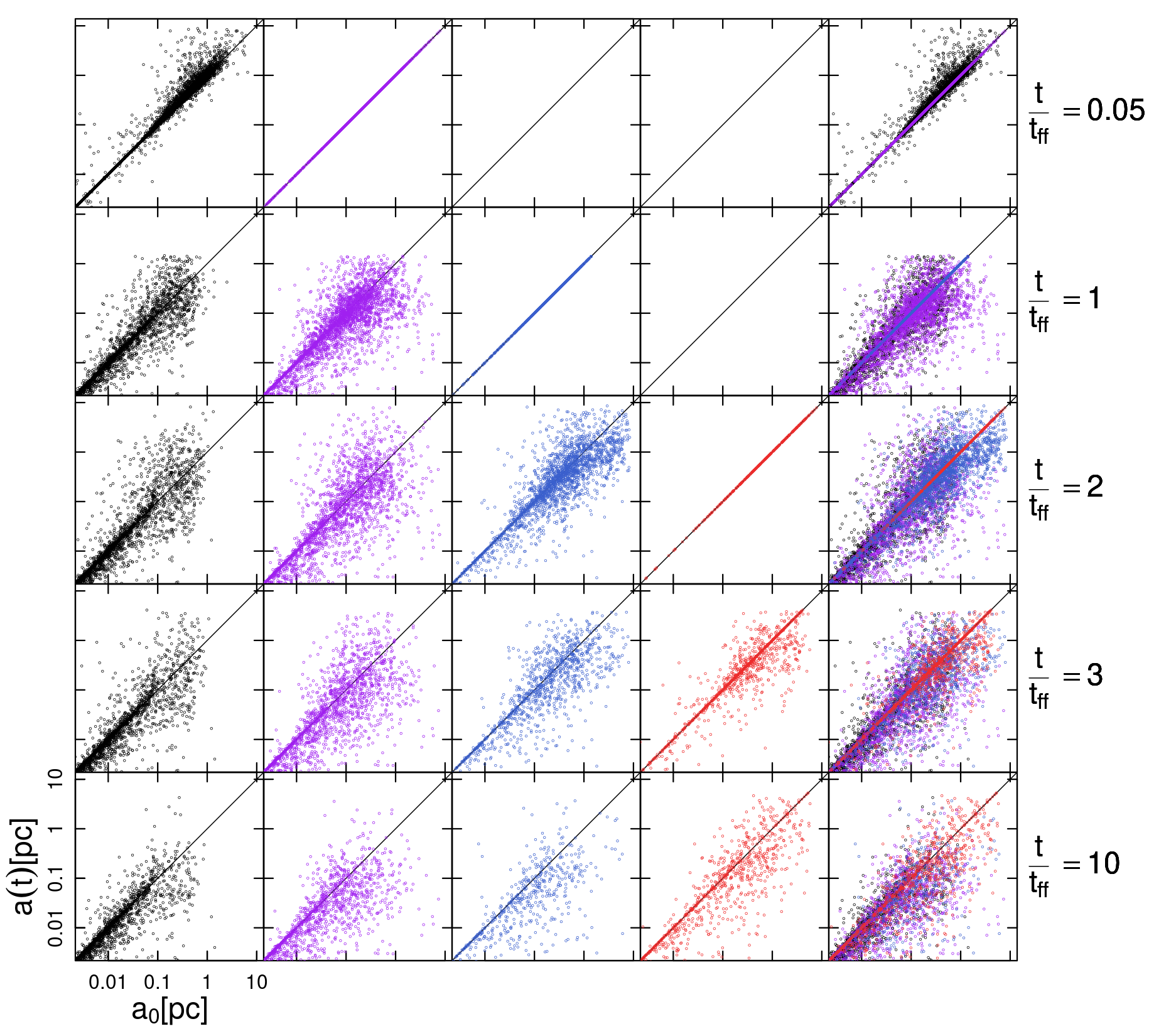}
    \caption{Same as Fig. \ref{fig:uvraa} for the \DVR\ model.}
    \label{fig:fvraa}
\end{figure*}

\begin{figure}
    \centering
    \includegraphics[width=0.4\textwidth]{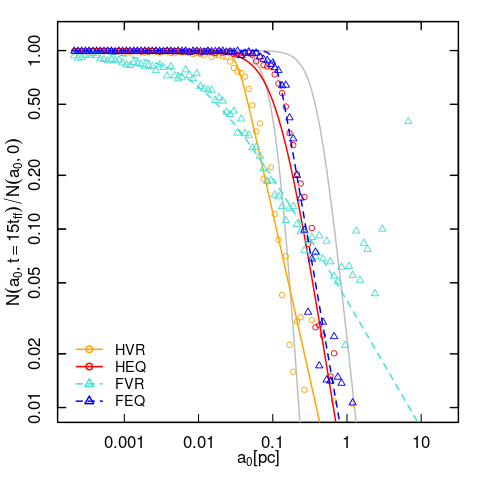}
    \includegraphics[width=0.4\textwidth]{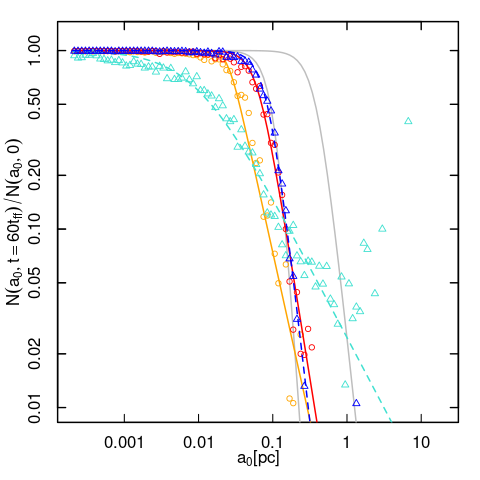}
    \caption{Fraction of surviving primordial binaries as a function of initial semi-major axis, $a_{\rm 0}$, after 15 $t/t_{\rm ff}$ (top) and 60 $t/t_{\rm ff}$ (bottom). The lines from the anisotropic (left grey line) and isotropic (right grey line) cusp models from \protect\cite{2016PeLu} are 
    plotted for reference. The violent relaxation models clearly show a more efficient disruption of intermediate width binaries relative to the reference anisotropic cusp model, while the two equilibrium models follow the profile of the same reference model.}
    \label{fig:intfracs}
\end{figure}
\begin{figure}
    \centering
    \includegraphics[width=0.4\textwidth]{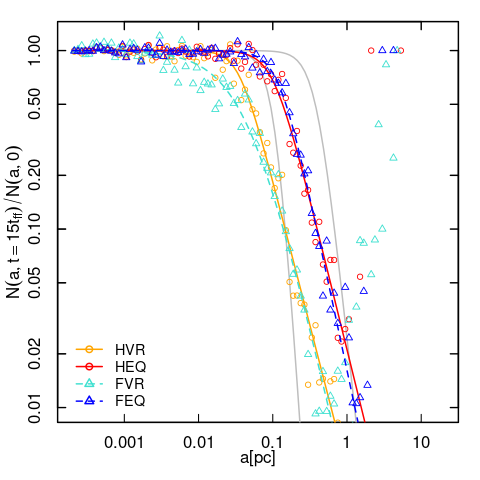}
    \includegraphics[width=0.4\textwidth]{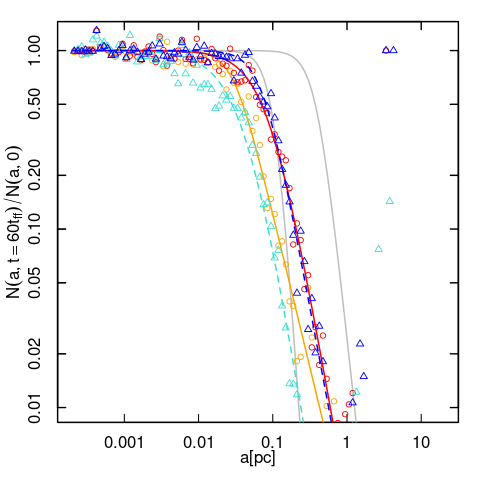}
    \caption{The distribution of the semi-major axis, $a$, of binaries after 15 $t/t_{\rm ff}$ (top) and 60 $t/t_{\rm ff}$ (bottom), normalized to the distribution of binaries at 0 $t/t_{\rm ff}$. The grey lines, plotted for reference, are the same as in Fig. \ref{fig:intfracs}. 
    }
    \label{fig:intfracstot}
\end{figure}

\begin{figure}
    \centering
    \includegraphics[width=0.4\textwidth]{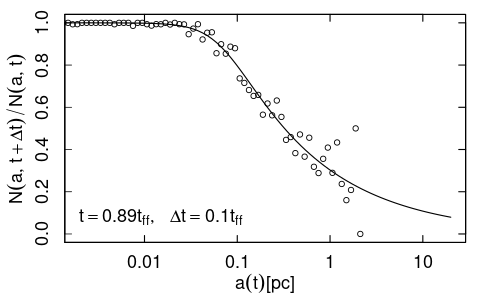}
    \includegraphics[width=0.4\textwidth]{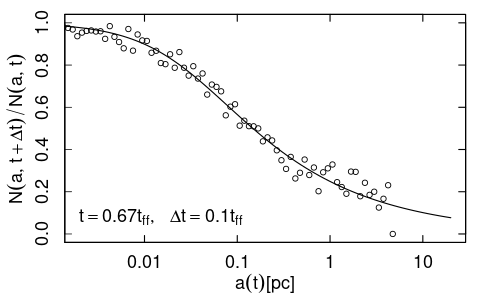}
    \includegraphics[width=0.4\textwidth]{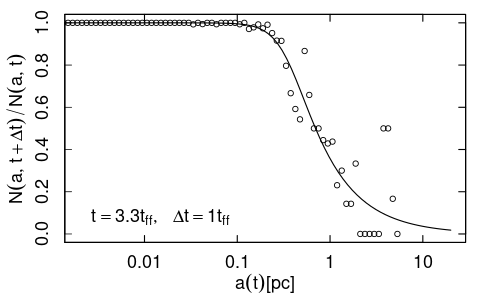}

    \caption{Best-fit survival curve (see equation \ref{eq:bpldt}) for the \UVR, \DVR, and \UEQ\ models (top to bottom) between $t$ and $t+\Delta t$ at a variety of times with binned data plotted as points for reference.}
    \label{fig:exfits}
\end{figure}

\begin{figure}
    \centering
    \includegraphics[width=0.45\textwidth]{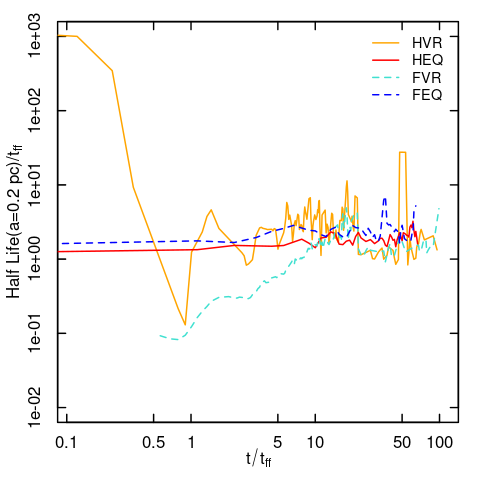}
    \caption{Time evolution of the half-life time of binary pairs with $a(t)=0.2$ pc. For all models, except the \DVR\ model, the half-life timescale is approximately constant after about 3 $t_{\rm ff}$. The collisional effects acting in the clumps surviving in the \DVR\ models leads to a more efficient binary disruption (shorter half-life timescale) until about 10 $t_{\rm ff}$.}
    \label{fig:halflives}
\end{figure}

\begin{figure}
    \centering
    \includegraphics[width=0.45\textwidth]{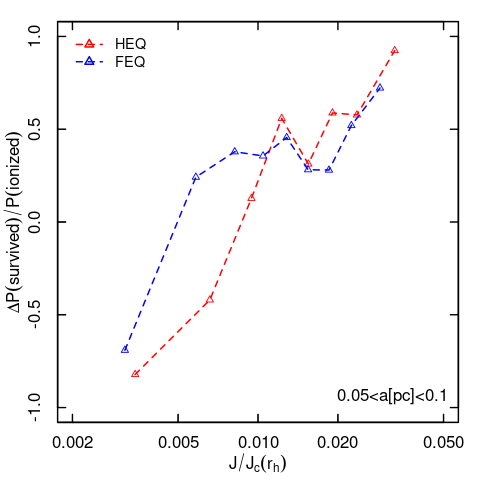}
    \includegraphics[width=0.45\textwidth]{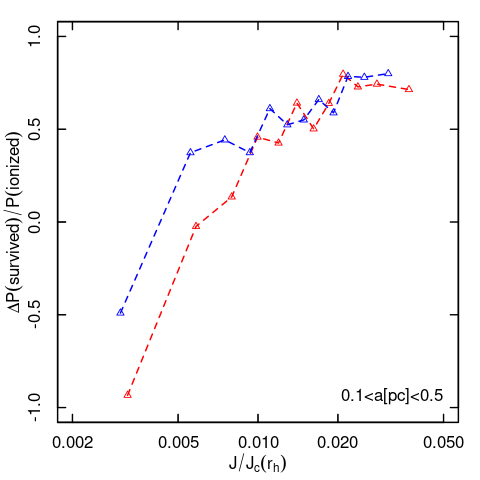}
  \caption{Residual of the survival probability (see equation \ref{eq:bpldt} for the probability model) versus the outer angular momentum of the orbit, $J$, for the two different semi-major axis ranges noted in the plots. The residual of the survival probability is normalized such that positive values imply an larger (for negative values, smaller) survival rate of binary pairs than expected. The angular momentum is normalized to the angular momentum of a circular orbit at the half-mass radius, $J_{\rm c}(r_{\rm h})$.}
    \label{fig:Jexp}
\end{figure}

\begin{figure*}
    \centering
    \includegraphics[width=0.4\textwidth]{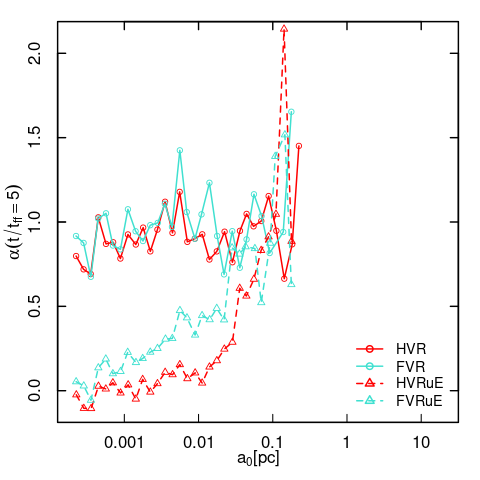}
    \includegraphics[width=0.4\textwidth]{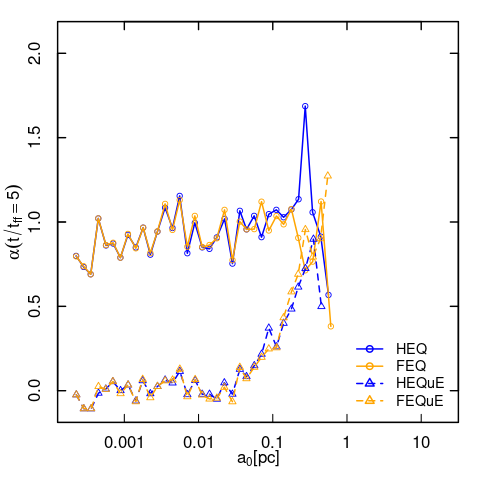}
    \includegraphics[width=0.4\textwidth]{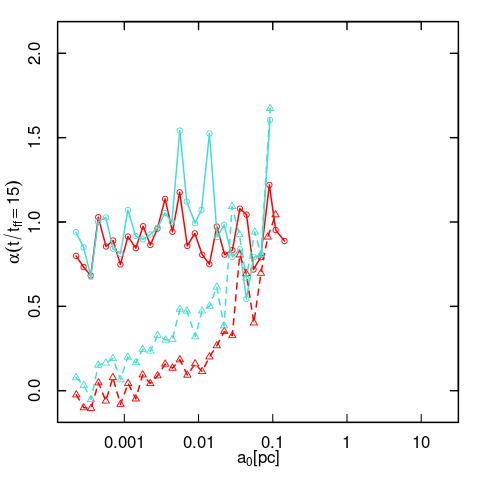}
    \includegraphics[width=0.4\textwidth]{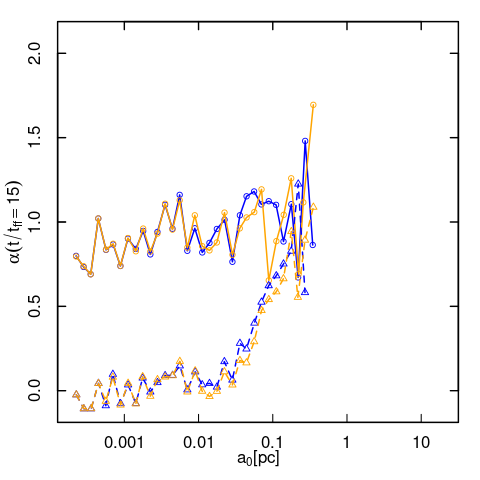}
  \caption{Power-law index of the eccentricity distribution, $\alpha$ (see equation \ref{eq:ecc}), as a function of $a_0$, at 5 $t_{\rm ff}$ (top) and 15 $t_{\rm ff}$ (bottom), for our violent relaxation models (left), and equilibrium models (right)}
    \label{fig:alphavsa}
\end{figure*}

\begin{figure*}
    \centering
    \includegraphics[width=0.4\textwidth]{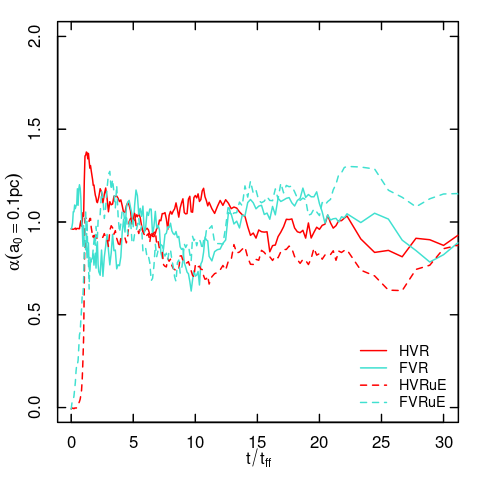}
    \includegraphics[width=0.4\textwidth]{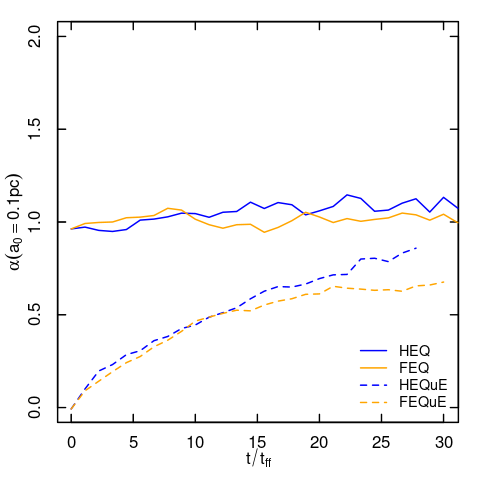}
   \caption{Time evolution of the index, $\alpha$ (see equation \ref{eq:ecc}), of the eccentricity distribution, for $a_0\approx 0.1$ pc in our violent relaxation models (left), and equilibrium models (right).}
    \label{fig:alphavst}
\end{figure*}

\subsection{Evolution and disruption of binaries in numerical simulations}
In this subsection, we present our results concerning the possible evolutionary paths leading to the disruption of binaries or to changes in their internal dynamical properties.

\subsubsection{Disruption of the widest binaries}
We start by focusing on binary disruption and show, in Figure \ref{fig:Bincount}, the time evolution of the fraction of surviving wide binary pairs (with initial semi-major axis, $a_0$, 0.2<$a_{0}$<2 pc) from set $\mathcal{I}$ in our models with an initially thermal eccentricity distribution. This plot clearly shows that during the violent relaxation phase, the binary disruption rate is significantly larger than that of systems starting in equilibrium and highlights the importance of considering this phase in the study of the evolution of a binary population. 

The discrepancy in the survival fraction between the violent relaxation and equilibrium models can be attributed to both the more extreme radial anisotropy within systems starting dynamically cold and undergoing violent relaxation as well as the collisional effects associated to stellar encounters during these early evolutionary phases.

In order to disentangle the effects associated with these two processes, we include in Figure \ref{fig:Bincount} the results of the \nColl\ models which follow the evolution of each individual binary in the dark matter potential. As discussed in Section \ref{sec:Methods}, these simulations include only the collisionless effects associated with the time variation of the external tidal field along each binary's orbit. 

While the time evolution of the fraction of wide binaries in the equilibrium system is similar between the $N$-body simulations and the collisionless integrations, they differ significantly in the case of the system undergoing violent relaxation. Specifically, this figure shows that the population of binaries in the full $N$-body simulation undergoes a stronger disruption than what is found in the collisionless integration: the additional binary disruption found in the full $N$-body simulation is a consequence of the effects associated with star-binary and binary-binary encounters occurring in the dense clumps characterizing the structure of the system during this early evolutionary phase and in the maximum contraction phase of the initial collapse.

\subsubsection{Evolution of the semi-major axis distribution}
In order to study the effects of the various dynamical processes affecting the properties of the binary population at early times, we start our analysis by comparing the values of the semi-major axes of binaries after 2.5 $t_{\rm ff}$, $a(2.5)$, with the corresponding initial semi-major axis of each binary, $a_0$, at the time each binary pair formed, $t_0$. By 2.5 $t_{\rm ff}$, the \VR\ models have undergone maximum contraction from the first free-fall and are evolving towards their final equilibrium configuration. We plot, in Figure \ref{fig:aarep}, $a(2.5)$ vs $a_0$ for the two \VR\ simulations and one of the equilibrium models (\UEQ).

This plot shows all binaries present at 2.5 $t_{\rm ff}$ including those in which, as a result of exchange interactions, at least one of the primordial components has been replaced by another star. 
This figure provides another clear manifestation of the strong collisional effects acting during the early evolutionary phases of the \VR\ simulations.
In particular the two panels for the \VR\ models show a more significant evolution of the binary semi-major axes and a much larger number of binaries undergoing exchange encounters. Although the initially wide binary pairs are destroyed over time, we can see that exchanges and widening of surviving binaries re-populate the wide semi-major axis tail of the distribution.

In order to further illustrate the complex evolution of binary stars during the early evolutionary stages of these systems, in Figures \ref{fig:uvraa} and \ref{fig:fvraa} we plot the values of the binary semi-major axes at different times versus their initial values for the \UVR\ and \DVR\ models, splitting the populations of new binaries formed in exchange encounters occurring at different times into different columns.

These two figures, in addition to showing the  evolution of the semi-major axes of primordial binaries, clearly demonstrate the role of collisional effects in producing new binaries and how the strength of these effects vary for the \UVR\ and the \DVR\ models.

In the \UVR\ model, new binary pairs form with predominantly very wide semi-major axes, and most of these pairs are ionized soon after, as shown by the lack of wide binaries in the panels corresponding to non-primordial binaries in the bottom row of Figure \ref{fig:uvraa}. 
The destruction of binaries by the end of the maximum contraction phase and the range of new binaries formed in the second free-fall time (shown in blue in the third row) both show how the stronger tidal field and the added collisionality during this phase play a significant role in the survival and mutation of binary pairs.

Due to its clumpy structure, the \DVR\ system is characterized by more complex dynamics.
As illustrated by the plots shown in Figures \ref{fig:uvraa} and \ref{fig:fvraa}, both a larger fraction of binaries survive after exchanges and a larger fraction of initially wide binaries survive the  early evolution of the stellar system in the \DVR\ model compared to the \UVR\ model, thanks to encounters causing their semi-major axes to shrink.
The prevalence of these effects come from the more frequent interactions within the stellar clumps.

Transitioning to the intermediate/long term evolution of binaries and the dependence of binary survival on semi-major axis, we plot the survival fraction of primordial binaries as a function of their initial semi-major axis, $a_0$, after 15 and 60 $t_{\rm ff}$ in Figure \ref{fig:intfracs}. These profiles provide further evidence of the more efficient binary disruption in the violent relaxation models; in particular, the violent relaxation models are significantly more efficient at destroying tighter binaries ($a_0\lesssim $0.1) than the equilibrium models.
The larger fraction of surviving wide binaries in the \DVR\ model is due to interactions leading to the shrinking of their semi-major axes resulting in tighter binaries.

In Figure \ref{fig:intfracstot}, we plot the final distribution of semi-major axis, $a$, after 15 and 60 $t_{\rm ff}$, normalized by the initial distribution of $a$. This complements Figure \ref{fig:intfracs} by showing how the evolution of the semi-major axes of the surviving binaries causes the final distribution of semi-major axes to be significantly different from that determined by the survival rate alone.

Following \cite{2016PeLu}, we fit our survival profiles with a broken power-law of the form:
\begin{equation}
    f(a)=\frac{1}{(1+(a/a_{\rm c})^m)^{n/m}}.
    \label{eq:bpl}
\end{equation}

The best fits to this function are overplotted as lines in Figures \ref{fig:intfracs} and \ref{fig:intfracstot}. This yields best-fit profiles (listed as [$a_{\rm c}$ (pc), $m$, $n$])  after 60 $t_{\rm ff}$ for our models in Figure \ref{fig:intfracs} as: [0.025, 6.3, 1.9] (\UVR), [0.0093, 1.1, 0.79] (\DVR), [0.062, 3.6, 2.6] (\UEQ), [0.086, 3.5, 3.7] (\DEQ), and in Figure \ref{fig:intfracstot} as: [0.030, 3.0, 1.7] (\UVR), [0.043, 1.6, 2.6] (\DVR), [0.088, 1.8, 2.4] (\UEQ), [0.073, 2.8, 2.2] (\DEQ).

After 60$t_{\rm ff}$ ($\approx$1 Gyr), the two equilibrium models, \UEQ\ and \DEQ, in both Figures \ref{fig:intfracs} and \ref{fig:intfracstot} roughly follow the anisotropic models used in \cite{2016PeLu}, which are measured after 10 Gyr, with all of our models show extra depletion for binary pairs near 0.1 pc. If our simulations are run to 10 Gyr, then we expect to see more depletion than our models at around 1 Gyr, as the evolution of binaries will continue to develop as the UFD ages further.

\subsection{Modeling binary survival}
\subsubsection{An evolving probability model}
Although this paper is focused on the effects of the early evolutionary stages and violent relaxation on the disruption of primordial binaries and the evolution of the properties of those that survive, we present here the results of a semi-analytical framework aimed at understanding and quantifying the disruption during different phases of the stellar system evolution.

In order to model the survival of binaries between individual snapshots separated by $\Delta t$, we model the decay of binaries as a probability based on an exponential decay function of equation \ref{eq:bpl}:
\begin{equation}
    P({\rm survival}|\ a,\Delta t)=\left[\frac{1}{(1+(a/a_{\rm c})^m)^{n/m}}\right]^{\Delta t/\tau}
    \label{eq:bpldt}
\end{equation}

\noindent where $\tau$ is a fixed timescale. We fit this model using the maximum likelihood function:
\begin{equation}
\mathcal{L}=\displaystyle\prod_{i=1}^N
\left\{\begin{array}{cc}
    P({\rm survival}|\ a_i,\Delta t)  & \rm{if\ survived} \\
    1-P({\rm survival}|\ a_i,\Delta t) & \rm{if\ ionized.}
\end{array}  \right.
\end{equation}

We show some example fits for intermediate times in Figure \ref{fig:exfits}. We can clearly see how the shape of this function changes significantly in the different models at different times. 
This model is flexible enough to properly model the binary survival both during the stellar system's collapse phase and the subsequent equilibrium phase. The only exception is the very early phases ($t<0.5t_{\rm ff}$) of the \DVR\ model, during which this expression is not adequate to describe the binary survival probability.

We plot the resulting half-life time for binary pairs with a representative semi-major axis of 0.2 pc as a function of time in Figure \ref{fig:halflives}. Data for the \DVR\ model in this figure starts at $t=0.5t_{\rm ff}$ because, as pointed out above, the expression in equation \ref{eq:bpldt} does not provide an adequate description of the binary survival probability during the very early evolution of this model.

\subsubsection{The effect of angular momentum}
While the semi-major axis is the primary parameter determining the evolution of a binary star in a stellar system, the outer orbit of the binary in the host system determines the extent of the time-variation and the strength of the tidal effects due to the external potential affecting the binary evolution. This subsection focuses on the study of the role of the binaries' orbits on the survival of binaries.

A comprehensive study of this issue would require a systematic and extensive exploration of the binaries' orbital phase space. Here, we provide a brief analysis illustrating the effect of only the binary's outer angular momentum on the binary's survival. As shown in Figure \ref{fig:EJstability}, for a given outer orbital energy, smaller values of the angular momentum imply small values of the minimal tidal radius achieved on a given orbit and directly affect the binary's survival.

In order to disentangle the effects of the semi-major axis and outer orbit, we use the following method: we measure the average residual of the survival fraction model fits (which are a function of $a$ as shown in Figure \ref{fig:exfits}) as a function of the outer angular momentum of the orbit, $J$. The result of this procedure is shown in Figure \ref{fig:Jexp}, where we plot the average residual probability gained or lost normalized by the chance of each binary pair being destroyed. These plots show a significantly modified probability of survival based on the outer orbit of the binary pair for binary ranges of 0.05-0.1 pc and 0.1-0.5 pc.

This trend further illustrates the interplay between the binary evolution/disruption and the binary outer orbit, and shows the consequences of binary disruption on the outer kinematic properties of the surviving binary population: binaries on more radial orbits are ionized more efficiently, leaving a population of surviving binaries characterized by a less radial orbital distribution. At the same time, the low angular momentum tail of the distribution may be in part replenished by more compact binaries on radial orbits that survive, but have their semi-majors axes widened by tidal effects.

\subsection{Evolution of the eccentricity distribution}
The distribution of binary eccentricity may provide key insights into the binary formation history and the dynamical processes affecting the evolution of the binary properties. 
A number of recent theoretical and observational investigations have focused on the binary eccentricity distribution and its evolution in different environments (see e.g. \citealt{2019GeLe}, \citealt{2020Toko}, \citealt{2023BeEv}, \citealt{2022HwTi}, \citealt{2022HwEl}, \citealt{2022HaCh}).

We devote this section to the study of the evolution of the binary eccentricity distribution during the violent relaxation and subsequent dynamical phases of our stellar systems.

In our simulations, we have explored a uniform initial eccentricity distribution and an initially thermal eccentricity distribution (a power-law distribution with index, $\alpha=1$; see equation \ref{eq:ecc}).

We visualize the dependence of $\alpha$ on the initial semi-major axis of our binary pairs in set $\mathcal{I}$ after 5 and 15  $t_{\rm ff}$ in Figure \ref{fig:alphavsa}.
The tight binaries are unaffected by the external tidal field and stellar encounters, and they preserve their initial eccentricity distribution. 
The internal parameters of wide binaries, on the other hand, undergo a significant evolution; for systems starting with a uniform eccentricity distribution in particular, the memory of this initial distribution quickly being erased for binaries of width $\gtrsim10^{-3}$ pc as the population of surviving binaries evolve towards a thermal distribution ($\alpha=1$) or superthermal ($\alpha>1$) distribution for the widest binaries. 
The results for models with an initial thermal distribution show that this distribution is more stable, and, in this case, only the wide semi-major axis tail of the distribution is characterized by a possible deviation from the initial thermal distribution.

The comparison between the evolution of the eccentricity distribution in the \VR\ and \EQ\ models provides further evidence of the dynamical effects associated with the violent relaxation phase and the presence of a clumpy structure in the early evolutionary phases of these systems.
Specifically, the four panels of Figure \ref{fig:alphavsa} show that the evolution away from a initial uniform eccentricity distribution is more significant  and affects binaries with smaller initial semi-major axes in the \VR\ models than the \EQ\ models. Moreover, the presence of clumps and substructure in the \DVR\ model and the additional effects of stellar encounters within these clumps causes a stronger evolution in this model compared to the \UVR\ model.

Finally, Figure \ref{fig:alphavst} shows the time evolution of the index $\alpha$ for binaries with $a_0=0.1$ pc and illustrates the rapid evolution towards a thermal/super-thermal eccentricity distribution during the very early evolutionary phases and collapse of the \VR\ models.

\section{Conclusions}
\label{sec:Concl}
In this paper, we have run a set of $N$-body simulations to explore the evolution of binary stars in ultra-faint dwarf galaxies. We have focused our attention on the early evolution of these systems, assuming that the stellar component of the UFD is initially dynamically cold and undergoes the typical collapse and structural oscillations of the violent relaxation phase.

Our simulations show that the UFD early dynamics and structural properties significantly affect the disruption and evolution of primordial binaries and leave their dynamical fingerprints on the properties of the population of surviving binaries.

As a result of the early collapse of the dynamically cold stellar component of the UFD, a large fraction of binaries move on low angular momentum (radial) orbits and wide binaries are disrupted by the tidal effects due to the potential of the UFD dark matter halo. Binary disruption and the evolution of the internal dynamical properties (semi-major axis and eccentricity) of the surviving binaries during this phase are enhanced compared to those found when the UFD is in equilibrium (see Figures \ref{fig:EJstability} and \ref{fig:Bincount}).

While collisional effects play a key role in the evolution of binary stars in star clusters (and, more in general, in their dynamical evolution), they are usually unimportant and neglected in the dynamics of  dwarf galaxies. In our study, we show these effects to be important during the high-density maximum contraction phase of the initial collapse and in the various clumps that may characterize the early structure of the systems we have explored before they reach a monolithic equilibrium structure. Our simulations show that these collisional effects associated to binary encounters with other single and binary stars give an additional significant contribution to the evolution and ionization of primordial binary stars (see Figures \ref{fig:Bincount}-\ref{fig:fvraa}). Stellar encounters also lead to a number of exchange interactions in which one of the original binary component is replaced by one of the interacting stars.

We have explored the evolution of the semi-major axis distribution and shown that models including the early violent relaxation phases are characterized by a stronger depletion of wide binaries than those starting in equilibrium, especially for binary pairs that are unlikely to be ionized by the tidal field alone (see Figures \ref{fig:intfracs} and \ref{fig:intfracstot}). Our analysis includes a semi-analytical description of the binary disruption  and its dependence on the binary semi-major axis and outer orbit in the host UFD.

Finally we have explored the evolution of the binary eccentricity distribution. In models starting with a thermal eccentricity distribution (power-law distribution with index, $\alpha=1$ see Equation \ref{eq:ecc}), this distribution is preserved during the early evolutionary phases except for the widest binaries for which the eccentricity distribution evolves towards and may display a super-thermal distribution.

 We have also studied models starting with a uniform eccentricity distribution ($\alpha=0$); in this case,  this distribution is preserved only for the most compact binaries (initial semi-major < $10^{-3}$ pc). For wider binaries, the eccentricity distribution rapidly evolves away from the initial uniform distribution towards power-law distributions with $\alpha>0$ and approach a thermal ($\alpha=1$) or superthermal  ($\alpha>1$) distribution for the widest binaries (see Figures \ref{fig:alphavsa} and \ref{fig:alphavst}). This evolution is more rapid in models starting with clumpy initial conditions due to the stronger collisional effects acting in the clumps characterizing the early structural properties of these systems (see Figure \ref{fig:alphavsa}).
 
The results of our study clearly show that dynamical processes acting during the UFD early evolutionary phases play an important role in the evolution of a population of primordial binary stars.
In a future study, the investigation presented here will be extended to carry out a survey exploring the combined effects of these early evolutionary phases with those due to different structural properties of the stellar system and its host dark matter halo. This comprehensive investigation will enable a closer comparison with observational investigations and allow to shed light on the role of these different factors in determining the present-day properties of binary stars in UFDs.

\section*{Acknowledgements}
This research was supported in part by Lilly Endowment, Inc., through its support for the Indiana University Pervasive Technology Institute.

\section*{DATA AVAILABILITY STATEMENT}
The data presented in this article may be shared on reasonable request to the corresponding author.

\bibliographystyle{mnras}
\bibliography{bib}

\begin{thebibliography}{}
\makeatletter
\relax
\def\mn@urlcharsother{\let\do\@makeother \do\$\do\&\do\#\do\^\do\_\do\%\do\~}
\def\mn@doi{\begingroup\mn@urlcharsother \@ifnextchar [ {\mn@doi@}
  {\mn@doi@[]}}
\def\mn@doi@[#1]#2{\def\@tempa{#1}\ifx\@tempa\@empty \href
  {http://dx.doi.org/#2} {doi:#2}\else \href {http://dx.doi.org/#2} {#1}\fi
  \endgroup}
\def\mn@eprint#1#2{\mn@eprint@#1:#2::\@nil}
\def\mn@eprint@arXiv#1{\href {http://arxiv.org/abs/#1} {{\tt arXiv:#1}}}
\def\mn@eprint@dblp#1{\href {http://dblp.uni-trier.de/rec/bibtex/#1.xml}
  {dblp:#1}}
\def\mn@eprint@#1:#2:#3:#4\@nil{\def\@tempa {#1}\def\@tempb {#2}\def\@tempc
  {#3}\ifx \@tempc \@empty \let \@tempc \@tempb \let \@tempb \@tempa \fi \ifx
  \@tempb \@empty \def\@tempb {arXiv}\fi \@ifundefined
  {mn@eprint@\@tempb}{\@tempb:\@tempc}{\expandafter \expandafter \csname
  mn@eprint@\@tempb\endcsname \expandafter{\@tempc}}}

\bibitem[\protect\citeauthoryear{{Benisty}, {Evans}  \& {Davis}}{{Benisty}
  et~al.}{2023}]{2023BeEv}
{Benisty} D.,  {Evans} N.~W.,   {Davis} A.-C.,  2023, \mn@doi [\mnras]
  {10.1093/mnrasl/slac134}, \href
  {https://ui.adsabs.harvard.edu/abs/2023MNRAS.518L..51B} {518, L51}

\bibitem[\protect\citeauthoryear{{Casertano} \& {Hut}}{{Casertano} \&
  {Hut}}{1985}]{CasHut}
{Casertano} S.,  {Hut} P.,  1985, \mn@doi [\apj] {10.1086/163589}, \href
  {https://ui.adsabs.harvard.edu/abs/1985ApJ...298...80C} {298, 80}

\bibitem[\protect\citeauthoryear{{Chandrasekhar}}{{Chandrasekhar}}{1944}]{1944Ch}
{Chandrasekhar} S.,  1944, \mn@doi [\apj] {10.1086/144589}, \href
  {https://ui.adsabs.harvard.edu/abs/1944ApJ....99...54C} {99, 54}

\bibitem[\protect\citeauthoryear{{El-Badry} \& {Rix}}{{El-Badry} \&
  {Rix}}{2018}]{2018ElRi}
{El-Badry} K.,  {Rix} H.-W.,  2018, \mn@doi [\mnras] {10.1093/mnras/sty2186},
  \href {https://ui.adsabs.harvard.edu/abs/2018MNRAS.480.4884E} {480, 4884}

\bibitem[\protect\citeauthoryear{{El-Badry}, {Rix}  \& {Heintz}}{{El-Badry}
  et~al.}{2021}]{2021ElRi}
{El-Badry} K.,  {Rix} H.-W.,   {Heintz} T.~M.,  2021, \mn@doi [\mnras]
  {10.1093/mnras/stab323}, \href
  {https://ui.adsabs.harvard.edu/abs/2021MNRAS.506.2269E} {506, 2269}

\bibitem[\protect\citeauthoryear{{Gaia Collaboration} et~al.,}{{Gaia
  Collaboration} et~al.}{2016}]{2016Gaia}
{Gaia Collaboration} et~al., 2016, \mn@doi [\aap]
  {10.1051/0004-6361/201629272}, \href
  {https://ui.adsabs.harvard.edu/abs/2016A&A...595A...1G} {595, A1}

\bibitem[\protect\citeauthoryear{{Geller}, {Leigh}, {Giersz}, {Kremer}  \&
  {Rasio}}{{Geller} et~al.}{2019}]{2019GeLe}
{Geller} A.~M.,  {Leigh} N. W.~C.,  {Giersz} M.,  {Kremer} K.,   {Rasio} F.~A.,
   2019, \mn@doi [\apj] {10.3847/1538-4357/ab0214}, \href
  {https://ui.adsabs.harvard.edu/abs/2019ApJ...872..165G} {872, 165}

\bibitem[\protect\citeauthoryear{{Hamilton}}{{Hamilton}}{2022}]{2022HaCh}
{Hamilton} C.,  2022, \mn@doi [\apjl] {10.3847/2041-8213/ac6600}, \href
  {https://ui.adsabs.harvard.edu/abs/2022ApJ...929L..29H} {929, L29}

\bibitem[\protect\citeauthoryear{{Hamilton} \& {Rafikov}}{{Hamilton} \&
  {Rafikov}}{2019a}]{2019HaRa}
{Hamilton} C.,  {Rafikov} R.~R.,  2019a, \mn@doi [\mnras]
  {10.1093/mnras/stz1730}, \href
  {https://ui.adsabs.harvard.edu/abs/2019MNRAS.488.5489H} {488, 5489}

\bibitem[\protect\citeauthoryear{{Hamilton} \& {Rafikov}}{{Hamilton} \&
  {Rafikov}}{2019b}]{2019HaRa2}
{Hamilton} C.,  {Rafikov} R.~R.,  2019b, \mn@doi [\mnras]
  {10.1093/mnras/stz2026}, \href
  {https://ui.adsabs.harvard.edu/abs/2019MNRAS.488.5512H} {488, 5512}

\bibitem[\protect\citeauthoryear{{Hartman} \& {L{\'e}pine}}{{Hartman} \&
  {L{\'e}pine}}{2020}]{2020HaLe}
{Hartman} Z.~D.,  {L{\'e}pine} S.,  2020, \mn@doi [\apjs]
  {10.3847/1538-4365/ab79a6}, \href
  {https://ui.adsabs.harvard.edu/abs/2020ApJS..247...66H} {247, 66}

\bibitem[\protect\citeauthoryear{{Heggie}}{{Heggie}}{1975}]{1975He}
{Heggie} D.~C.,  1975, \mn@doi [\mnras] {10.1093/mnras/173.3.729}, \href
  {https://ui.adsabs.harvard.edu/abs/1975MNRAS.173..729H} {173, 729}

\bibitem[\protect\citeauthoryear{{Hwang}, {Ting}  \& {Zakamska}}{{Hwang}
  et~al.}{2022a}]{2022HwTi}
{Hwang} H.-C.,  {Ting} Y.-S.,   {Zakamska} N.~L.,  2022a, \mn@doi [\mnras]
  {10.1093/mnras/stac675}, \href
  {https://ui.adsabs.harvard.edu/abs/2022MNRAS.512.3383H} {512, 3383}

\bibitem[\protect\citeauthoryear{{Hwang}, {El-Badry}, {Rix}, {Hamilton}, {Ting}
   \& {Zakamska}}{{Hwang} et~al.}{2022b}]{2022HwEl}
{Hwang} H.-C.,  {El-Badry} K.,  {Rix} H.-W.,  {Hamilton} C.,  {Ting} Y.-S.,
  {Zakamska} N.~L.,  2022b, \mn@doi [\apjl] {10.3847/2041-8213/ac7c70}, \href
  {https://ui.adsabs.harvard.edu/abs/2022ApJ...933L..32H} {933, L32}

\bibitem[\protect\citeauthoryear{{Jiang} \& {Tremaine}}{{Jiang} \&
  {Tremaine}}{2010}]{2010JiTr}
{Jiang} Y.-F.,  {Tremaine} S.,  2010, \mn@doi [\mnras]
  {10.1111/j.1365-2966.2009.15744.x}, \href
  {https://ui.adsabs.harvard.edu/abs/2010MNRAS.401..977J} {401, 977}

\bibitem[\protect\citeauthoryear{{Kervick}, {Walker}, {Pe{\~n}arrubia}  \&
  {Koposov}}{{Kervick} et~al.}{2022}]{2022KeWa}
{Kervick} C.,  {Walker} M.~G.,  {Pe{\~n}arrubia} J.,   {Koposov} S.~E.,  2022,
  \mn@doi [\apj] {10.3847/1538-4357/ac5b5f}, \href
  {https://ui.adsabs.harvard.edu/abs/2022ApJ...929...77K} {929, 77}

\bibitem[\protect\citeauthoryear{{K{\"u}pper}, {Maschberger}, {Kroupa}  \&
  {Baumgardt}}{{K{\"u}pper} et~al.}{2011}]{2011KuMa}
{K{\"u}pper} A. H.~W.,  {Maschberger} T.,  {Kroupa} P.,   {Baumgardt} H.,
  2011, \mn@doi [\mnras] {10.1111/j.1365-2966.2011.19412.x}, \href
  {https://ui.adsabs.harvard.edu/abs/2011MNRAS.417.2300K} {417, 2300}

\bibitem[\protect\citeauthoryear{{Lah{\'e}n}, {Naab}, {Johansson}, {Elmegreen},
  {Hu}  \& {Walch}}{{Lah{\'e}n} et~al.}{2020}]{2020LaNa}
{Lah{\'e}n} N.,  {Naab} T.,  {Johansson} P.~H.,  {Elmegreen} B.,  {Hu} C.-Y.,
  {Walch} S.,  2020, \mn@doi [\apj] {10.3847/1538-4357/abc001}, \href
  {https://ui.adsabs.harvard.edu/abs/2020ApJ...904...71L} {904, 71}

\bibitem[\protect\citeauthoryear{{McConnachie} \& {C{\^o}t{\'e}}}{{McConnachie}
  \& {C{\^o}t{\'e}}}{2010}]{2010McCo}
{McConnachie} A.~W.,  {C{\^o}t{\'e}} P.,  2010, \mn@doi [\apjl]
  {10.1088/2041-8205/722/2/L209}, \href
  {https://ui.adsabs.harvard.edu/abs/2010ApJ...722L.209M} {722, L209}

\bibitem[\protect\citeauthoryear{{Minor}, {Pace}, {Marshall}  \&
  {Strigari}}{{Minor} et~al.}{2019}]{2019MiPa}
{Minor} Q.~E.,  {Pace} A.~B.,  {Marshall} J.~L.,   {Strigari} L.~E.,  2019,
  \mn@doi [\mnras] {10.1093/mnras/stz1468}, \href
  {https://ui.adsabs.harvard.edu/abs/2019MNRAS.487.2961M} {487, 2961}

\bibitem[\protect\citeauthoryear{{Navarro}, {Frenk}  \& {White}}{{Navarro}
  et~al.}{1997}]{1997NFW}
{Navarro} J.~F.,  {Frenk} C.~S.,   {White} S. D.~M.,  1997, \mn@doi [\apj]
  {10.1086/304888}, \href
  {https://ui.adsabs.harvard.edu/abs/1997ApJ...490..493N} {490, 493}

\bibitem[\protect\citeauthoryear{{Oh}, {Price-Whelan}, {Hogg}, {Morton}  \&
  {Spergel}}{{Oh} et~al.}{2017}]{2017OhPr}
{Oh} S.,  {Price-Whelan} A.~M.,  {Hogg} D.~W.,  {Morton} T.~D.,   {Spergel}
  D.~N.,  2017, \mn@doi [\aj] {10.3847/1538-3881/aa6ffd}, \href
  {https://ui.adsabs.harvard.edu/abs/2017AJ....153..257O} {153, 257}

\bibitem[\protect\citeauthoryear{{Pe{\~n}arrubia}}{{Pe{\~n}arrubia}}{2021}]{2021PeJo}
{Pe{\~n}arrubia} J.,  2021, \mn@doi [\mnras] {10.1093/mnras/staa3700}, \href
  {https://ui.adsabs.harvard.edu/abs/2021MNRAS.501.3670P} {501, 3670}

\bibitem[\protect\citeauthoryear{{Pe{\~n}arrubia}, {Ludlow}, {Chanam{\'e}}  \&
  {Walker}}{{Pe{\~n}arrubia} et~al.}{2016}]{2016PeLu}
{Pe{\~n}arrubia} J.,  {Ludlow} A.~D.,  {Chanam{\'e}} J.,   {Walker} M.~G.,
  2016, \mn@doi [\mnras] {10.1093/mnrasl/slw090}, \href
  {https://ui.adsabs.harvard.edu/abs/2016MNRAS.461L..72P} {461, L72}

\bibitem[\protect\citeauthoryear{{Ramirez} \& {Buckley}}{{Ramirez} \&
  {Buckley}}{2022}]{2022RaBu}
{Ramirez} E.~D.,  {Buckley} M.~R.,  2022, arXiv e-prints, \href
  {https://ui.adsabs.harvard.edu/abs/2022arXiv220908100R} {p. arXiv:2209.08100}

\bibitem[\protect\citeauthoryear{{Ricotti}, {Parry}  \& {Gnedin}}{{Ricotti}
  et~al.}{2016}]{2016RiPa}
{Ricotti} M.,  {Parry} O.~H.,   {Gnedin} N.~Y.,  2016, \mn@doi [\apj]
  {10.3847/0004-637X/831/2/204}, \href
  {https://ui.adsabs.harvard.edu/abs/2016ApJ...831..204R} {831, 204}

\bibitem[\protect\citeauthoryear{{Safarzadeh}, {Simon}  \& {Loeb}}{{Safarzadeh}
  et~al.}{2022}]{2022SaSi}
{Safarzadeh} M.,  {Simon} J.~D.,   {Loeb} A.,  2022, \mn@doi [\apj]
  {10.3847/1538-4357/ac626e}, \href
  {https://ui.adsabs.harvard.edu/abs/2022ApJ...930...54S} {930, 54}

\bibitem[\protect\citeauthoryear{{Simon}}{{Simon}}{2019}]{2019Si}
{Simon} J.~D.,  2019, \mn@doi [\araa] {10.1146/annurev-astro-091918-104453},
  \href {https://ui.adsabs.harvard.edu/abs/2019ARA&A..57..375S} {57, 375}

\bibitem[\protect\citeauthoryear{{Spencer}, {Mateo}, {Walker}, {Olszewski},
  {McConnachie}, {Kirby}  \& {Koch}}{{Spencer} et~al.}{2017}]{2017SpMa}
{Spencer} M.~E.,  {Mateo} M.,  {Walker} M.~G.,  {Olszewski} E.~W.,
  {McConnachie} A.~W.,  {Kirby} E.~N.,   {Koch} A.,  2017, \mn@doi [\aj]
  {10.3847/1538-3881/aa6d51}, \href
  {https://ui.adsabs.harvard.edu/abs/2017AJ....153..254S} {153, 254}

\bibitem[\protect\citeauthoryear{{Tian}, {El-Badry}, {Rix}  \& {Gould}}{{Tian}
  et~al.}{2020}]{2020TiEl}
{Tian} H.-J.,  {El-Badry} K.,  {Rix} H.-W.,   {Gould} A.,  2020, \mn@doi
  [\apjs] {10.3847/1538-4365/ab54c4}, \href
  {https://ui.adsabs.harvard.edu/abs/2020ApJS..246....4T} {246, 4}

\bibitem[\protect\citeauthoryear{{Tokovinin}}{{Tokovinin}}{2020}]{2020Toko}
{Tokovinin} A.,  2020, \mn@doi [\mnras] {10.1093/mnras/staa1639}, \href
  {https://ui.adsabs.harvard.edu/abs/2020MNRAS.496..987T} {496, 987}

\bibitem[\protect\citeauthoryear{{Tokovinin} \& {Kiyaeva}}{{Tokovinin} \&
  {Kiyaeva}}{2016}]{2016ToKi}
{Tokovinin} A.,  {Kiyaeva} O.,  2016, \mn@doi [\mnras] {10.1093/mnras/stv2825},
  \href {https://ui.adsabs.harvard.edu/abs/2016MNRAS.456.2070T} {456, 2070}

\bibitem[\protect\citeauthoryear{{Wang}, {Spurzem}, {Aarseth}, {Nitadori},
  {Berczik}, {Kouwenhoven}  \& {Naab}}{{Wang} et~al.}{2015}]{2015WaRa}
{Wang} L.,  {Spurzem} R.,  {Aarseth} S.,  {Nitadori} K.,  {Berczik} P.,
  {Kouwenhoven} M.~B.~N.,   {Naab} T.,  2015, \mn@doi [\mnras]
  {10.1093/mnras/stv817}, \href
  {https://ui.adsabs.harvard.edu/abs/2015MNRAS.450.4070W} {450, 4070}

\bibitem[\protect\citeauthoryear{{Weinberg}, {Shapiro}  \&
  {Wasserman}}{{Weinberg} et~al.}{1987}]{1987WeSh}
{Weinberg} M.~D.,  {Shapiro} S.~L.,   {Wasserman} I.,  1987, \mn@doi [\apj]
  {10.1086/164883}, \href
  {https://ui.adsabs.harvard.edu/abs/1987ApJ...312..367W} {312, 367}

\makeatother
\end{thebibliography}
\end{document}